\documentclass[twocolumn,english,showpacs,floatfix,aps,pra,eqsecnum]{revtex4}
\usepackage[T1]{fontenc}
\usepackage[latin1]{inputenc}
\usepackage{color}
\usepackage{graphicx}

\makeatletter

\usepackage{amssymb}
\usepackage{amsmath}

\usepackage{dcolumn} 
\usepackage{bm} 

\usepackage{babel}

\makeatother
\begin{document}

\title{Symmetry breaking effects upon bipartite and multipartite entanglement
in the XY model}

\author{Thiago R. de Oliveira}

\email{tro@ifi.unicamp.br}

\author{Gustavo Rigolin}

\email{rigolin@ifi.unicamp.br}

\author{Marcos C. de Oliveira}

\email{marcos@ifi.unicamp.br}

\author{E. Miranda}

\email{emiranda@ifi.unicamp.br}

\affiliation{Instituto de F\'{\i}sica Gleb Wataghin, Universidade Estadual de Campinas,
Caixa Postal 6165, CEP 13083-970, Campinas, SP, Brazil}

\begin{abstract}
We analyze the bipartite and multipartite entanglement for the ground
state of the one-dimensional XY model in a transverse magnetic field
in the thermodynamical limit. We explicitly take into account the
spontaneous symmetry breaking in order to explore the relation between
entanglement and quantum phase transitions. As a result we show that
while both bipartite and multipartite entanglement can be enhanced
by spontaneous symmetry breaking deep into the ferromagnetic phase,
only the latter is affected by it in the vicinity of the critical
point. This result adds to the evidence that multipartite, and not
bipartite, entanglement is the fundamental indicator of long range
correlations in quantum phase transitions. 
\end{abstract}


\maketitle

\section{Introduction}

One of the most remarkable features of many phase transitions is the
occurrence of spontaneous symmetry breaking, in which the symmetry of
the Hamiltonian is not realized in the system state. When this occurs
a macroscopic observable (the order parameter) emerges, which is
required for a unique specification of the microscopic state
\cite{and1}.  Those quantum phase transitions (QPT) occur at zero
temperature, and are triggered by the variation of a parameter of the
system's Hamiltonian \cite{sachdev}. The system's eigenenergies then
show non-analytical behavior which embodies the order of the phase
transition. These non-analyticities in turn are reflected in several
macroscopic observables.

Lately there has been an increasing interest in describing QPTs, not
by means of the non-analyticities of the spectrum or of the physical
observables, but rather by the amount of entanglement (bipartite or
multipartite) present in each of the system's phases. This is
motivated in particular by a general expectation that entanglement can
be given as important a status as the energy: both quantities can be
seen as resources useful for the accomplishment of interesting
physical tasks \cite{livronielsen}. Moreover, long-range correlations
often found in strongly correlated many-body systems at zero
temperature have a purely quantum origin and are expected to be
inextricable from entanglement \cite{zanardi,tobias}. Several authors
have studied the role of entanglement in QPTs by considering either
bipartite or multipartite entanglement measures (e.g., see
\cite{nielsen}-\cite{latorre RG}), both calculated for spin-$1/2$
lattice models such as the Ising and XY model in a transverse magnetic
field \cite{Ising original,McCoy1,McCoy2,McCoy3}.  Invariably though
many of these developments do not take into account the spontaneous
symmetry breaking accompanying the QPT, employing instead symmetric
states in calculations in the ordered phase (for an exception 
see \cite{dinamarques,Osterloh} and the note at the end of the
manuscript). This procedure, although
common and even correct for finite systems, is unrealistic in the
study of QPTs since it is well known that symmetric states
(Schr\"{o}dinger cats) are never realized in the thermodynamic limit
due to superselection rules \cite{and1}.

Recently, we have developed a program on the investigation of
multipartite entanglement properties in QPTs by proposing a
Generalized Global Entanglement (GGE) measure \cite{nosso paper
  1,newmeasure}. As the name suggests this is a generalization of the
Meyer and Wallach Global Entanglement \cite{meyer} (here dubbed
$G(1)$) to account for all the possible bipartitions of the system
state. Besides being able to detect any kind of entanglement present
in the system \cite{newmeasure}, this measure is also able to signal
the location and the type (order) of QPTs \cite{nosso paper 3}.  Our
results developed for the infinite 1-D Ising and XY chains indicate
that multipartite entanglement is maximal at the critical point,
playing a major role in the QPT process, in contrast to bipartite
entanglement \cite{nielsen,Nature}.

To correctly consider the spontaneous symmetry breaking due to quantum
fluctuations is an essential element for the success of such an
entanglement measure. However this point is particularly unclear in
the recent literature on entanglement in QPTs. Many authors are not
specific in their choices and have erroneously applied a symmetric
state in their investigations. In this article we extend the
entanglement analysis for the XY model taking into account the role
played by spontaneous symmetry breaking in both bipartite (pairwise)
and multipartite entanglement.  As we argue in the body of the paper,
symmetry breaking favors multipartite entanglement.  While bipartite
measures (concurrence and negativity) are the same irrespective of the
state one is employing, multipartite entanglement is not. We show that
both $G(1)$ as well as $G(2,n)$, an auxiliary function defining one
class of GGE, exhibit completely different behaviors depending on
whether the state is symmetric or not. We also review some of the
results on bipartite and multipartite entanglement in a clearer and
more detailed fashion. We expect with this work to settle some issues
concerning the relation between entanglement and quantum phase
transitions, at least in the context of the one-dimensional XY model.
This paper is structured as follows. In Sec. \ref{sec:entqpt} we
review the essential results in the literature regarding bipartite and
multipartite entanglement present in the 1-D Ising and XY models. In
Sec.~\ref{sec:XY-model} we discuss the XY model in detail and show how
the reduced two-spin state is calculated.  In
Sec.~\ref{sec:Bipartite-Entanglement} we analyze bipartite
entanglement measures (concurrence and negativity) by comparing the
results obtained for the symmetric state with the ones obtained for
the broken-symmetry one. In Sec.~\ref{sec:Multipartite-Entanglement}
we analyze the multipartite entanglement as given by $G(1)$ and
$G(2,1)$ for the two choices of ground states (symmetric or
broken-symmetry). Finally, in Sec.~\ref{sec:Conclusions} a discussion
ends the paper.

\section{Entanglement and QPT in 1-D Ising and XY spin chains}\label{sec:entqpt}

In this section we outline some of the most relevant findings \cite{footnote} associated
with entanglement in the 1-D Ising and XY models.

The first approach we mention, concerning pairwise entanglement (concurrence)
between two spins in the chain, was considered in Refs. \cite{nielsen,Nature}.
It was demonstrated that the concurrence between nearest neighbors
of the XY model is maximal not at the critical point but in its vicinity.
Furthermore, the pairwise entanglement between neighbors more than
three sites apart vanishes in the quantum Ising chain. The authors
of Ref. \cite{Nature} also showed that the derivative of the nearest
neighbors concurrence is able to signal the QPT as it diverges at the
critical point and exhibits finite-size scaling. Thus, it became clear
that the ability to signal a QPT could be a general property of good
entanglement measures.

The first work to establish a formal relation between a QPT and bipartite
entanglement measures was Ref. \cite{sarandy}. The authors have demonstrated
that, under a set of reasonable assumptions, a discontinuity in a
\textit{bipartite} entanglement measure (concurrence \cite{wootters}
and negativity \cite{werner}) is a necessary and sufficient indicator
of a first order quantum phase transition (1QPT), which is generically
characterized by a discontinuity in the first derivative of the ground
state energy. Furthermore, they have shown that a discontinuity or
a divergence in the first derivative of the same measure (assuming
it is continuous) is a necessary and sufficient indicator of a second
order QPT (2QPT), which is generically characterized by a discontinuity
or a divergence of the second derivative of the ground state energy.
Subsequently, it was pointed out \cite{venuti} that this result was
more general and would apply to any entanglement measure dependent
on the reduced density operator of two spins. Finally, it was demonstrated
in Ref. \cite{sarandy2}, using the Density Functional Theory formalism,
that any entanglement measure can be expressed as a unique functional
of the set of first derivatives of the ground state energy. For most
of the cases, however, the explicit expression of the functional is
not known. This result showed that any entanglement measure can in
principle signal a QPT, since it inherits the non-analytical behavior
of the derivative of the energy. Of course, depending on the definition
of the entanglement measure used, {}``accidental'' cancellations
of such divergences/discontinuities may occur (see Refs. \cite{yang,j vidal}).
Another approach to understand pairwise entanglement in QPTs based
on the study of the crossing of energy levels has also been proposed
\cite{shi-jian2}. For the case of a system of indistinguishable particles
it was proved that, given some provisos, the entanglement between
one part (A) and the rest (B) is able to signal a QPT \cite{larsson}.
However, in this case parts A and B correspond to modes not particles,
in contrast to the former mentioned works. We should also note that
pairwise entanglement in small chains (two, three, and four spins)
for the XY model was previously studied in Ref. \cite{wang}.

The second kind of approach worth mentioning focuses on multipartite
entanglement (ME). In Refs. \cite{latorre1,latorre2} the entropy of
entanglement between one part of the chain (a block of $L$ spins) and
the rest is employed for this purpose. There the entanglement entropy
is defined through the von Neumann entropy of one of the reduced
parts, a valid approach whenever the global state is pure. It was
shown for some spin-1/2 models that at the critical point (CP) the
entanglement entropy increases logarithmically with $L$, whereas it
saturates for large $L$ away from the CP, a result which had been
known from conformal field theory \cite{holzhey,cardy}.  For the
one-dimensional XY model the block entanglement was extensively and
carefully studied in Ref. \cite{korepin}. Another approach for ME
investigation (ME) is considered in Ref. \cite{wei} through the study
of the maximal possible overlap between the state studied and all
possible separable states; the larger this overlap is, the less
entangled the state. The XY model was analyzed in that way and it was
shown that the ME is maximal around the CP and its derivatives diverge
as the CP is approached. It was also shown that the ME is zero at the
second critical point (2CP) where the state is known to be separable
\cite{tognetti} (see the discussion in
Sec.~\ref{sec:Bipartite-Entanglement}).  Tripartite entanglement,
given in terms of the residual tangle \cite{coffman}, was also
analyzed for an Ising chain of 3 spins in a transverse field in Ref.
\cite{buzek}. In this last work it was shown that the residual tangle
is not maximal around the expected CP, which really exists only in the
thermodynamic limit. With the purpose of studying ME, a new measure
was defined and analyzed for spin chains in Ref. \cite{LE}.  It was
named Localizable Entanglement and defined as the maximal amount of
entanglement that can be localized in two particles, on average, by
doing local measurements on the rest of the particles. The Localizable
Entanglement was shown to be maximal at the {}``critical'' point for a
finite Ising chain of 14 spins. This, together with the results for
the block entanglement, were the first evidences that multipartite
entanglement could be important in the context of quantum phase
transitions.  It was also demonstrated that connected correlation
functions are a lower bound for Localizable Entanglement, a remarkable
result enabling the system to have a finite correlation length but
infinite entanglement length (see \cite{LE2} for an example of such
behavior).  We should remark, however, that it has been argued that
Localizable Entanglement may not be an entanglement monotone
\cite{LE3}.

With the hope that the tools of Quantum Information and Computation
could help to better understand Quantum Phase transitions Zannardi
and coworkers \cite{fidelity} have proposed and (in their own words)
{}``showed that quantum fidelity - the overlap modulus - of two finite-size
ground states corresponding to neighboring control parameters is a
good indicator of quantum phase transitions. Indeed, the fidelity
typically drops abruptly at the critical points, as a consequence
of the dramatic state transformation involved in a transition''.
For the sake of completeness we should also mention studies of the
temporal evolution of the entanglement in the XY chain \cite{time evolution}
as well as other attempts to show that multipartite entanglement is
important/enhanced in quantum phase transitions, as can be found in
references \cite{tognetti,tommaso,korepin2,yan1,yan2}.

We note that none of the employed entanglement measures in the above
studies are maximal at the CP, with the exception of the single site
entropy of the Ising model \cite{nielsen} in the thermodynamic limit
and the Localizable Entanglement \cite{LE} of an Ising chain of a
few spins. We should also mention that in Ref. \cite{latorre RG}
the authors have studied the loss of entanglement along the renormalization
group flow of an XY chain. For this purpose they obtained the entanglement
between 100 spins and the rest of the chain, as a function of the
transverse magnetic field and the anisotropy, showing that it was
indeed maximal at the critical point. 

At this point we should mention another interesting feature observed
in Refs. \cite{nielsen,Nature}, independently. They showed that bipartite
entanglement vanishes when the distance between the two spins is greater
than one lattice site. This is quite surprising since long range quantum
correlations are expected to be present at the CP. It was then conjectured
that bipartite entanglement at the CP would decrease in order for
the ME to increase, due to entanglement sharing \cite{nielsen}. In
other words, ME only appears at the expense of pairwise entanglement
and at the CP we should expect a genuine multipartite entangled state.

In Refs. \cite{nosso paper 1,newmeasure} three of us used the fact
that for the quantum Ising chain the entanglement between one spin
and the rest of the chain (given by the averaged linear entropy \cite{brennen}) is equal to the Global
Entanglement (GE), $G(1)$, a proposed ME measure introduced by Meyer and
Wallach in 2002 \cite{meyer}, in order to show that $G(1)$ is maximal
at the critical point. Inspired by the GE, we have also proposed the GGE, $E_{G}^{(n)}$, where the averages
are taken over the linear entropy of two, three, and more spins (or
subsystems). A similar approach was independently presented in Ref.~\cite{scott}.
In that construction, we allow the spins to be non-contiguous along the
chain and not just in a continuous block as already considered in
Refs. \cite{latorre1,latorre2}. For example, $E_{G}^{(2)}$ is the
entanglement between two spins and the rest of the chain averaged
over all possible distances between the spins. In this context, another
quantity that is also interesting is the average entanglement between
two spins $n$ sites apart and the rest of the chain (without averaging
over $n$): $G(2,n)$. In Refs. \cite{nosso paper 1,newmeasure} we then analyzed
the entanglement between two spins $n$ sites apart ($G(2,n)$) and
the rest of the chain showing that it is maximal at the critical point
and increases with $n$, saturating at the value 0.675 in the limit
of large $n$. This result was one of the first indications that multipartite
entanglement is maximal/enhanced and more distributed at the critical
point, adding strength to the conjecture of T. J. Osborne and M. A.
Nielsen \cite{nielsen}. It also suggested that ME is the key ingredient
for the appearance of the long-range correlations that develop at
the critical point. In Ref. \cite{newmeasure} we have pursued the
discussion of Ref. \cite{nosso paper 1} further, exploring the advantageous
features of the Generalized Global Entanglement for an operational
multipartite entanglement classification and quantification, in comparison
to the other available measures for both finite and infinite collections
of two-level systems.

In a more recent work \cite{nosso paper 3}, we extended the above
results to the one-dimensional XY model showing that $G(1)$ and $G(2,n)$
are maximal at the critical point. In a more general context we also
showed explicitly that $G(2,n)$ is able to signal QPTs, something
already expected from the results of \cite{venuti,sarandy2}. Finally,
and more striking, we demonstrated that for collections of two-level
systems with symmetry-breaking second-order quantum phase transitions,
$G(2,n)$ increases exponentially with $n$ away from the critical
point. This increase is governed by a characteristic length, named
the entanglement length $\xi_{E}$, which is half the correlation
length $\xi_{E}=\xi_{C}/2$. Furthermore, at the critical point $G(2,n)$
increases as a power law, implying an infinite entanglement length.
In fact, $\xi_{E}$ inherits the full critical behavior of $\xi_{C}$,
with the same critical exponent. All these results indicate again
that multipartite entanglement plays a major role at quantum phase
transitions, as argued in the previous paragraph.
How important is the spontaneous symmetry breaking for this conclusion? In the next sections we develop this question for the 1-D XY spin chain.

\section{XY model and the two-spin reduced density matrix}

\label{sec:XY-model}

The one-dimensional XY model in a transverse field is governed by
the following Hamiltonian \begin{equation}
H=-\sum_{i=1}^{N}\frac{J}{2}[(1+\gamma)\sigma_{i}^{x}\sigma_{i+1}^{x}+(1-\gamma)\sigma_{i}^{y}\sigma_{i+1}^{y}]+h\sum_{i=1}^{N}\sigma_{i}^{z},\label{xymodel}\end{equation}
 where $\sigma_{i}^{\alpha}$, $\alpha=x,y,z$, are the usual Pauli
matrices. The model reduces to the quantum Ising model for $\gamma=1$
and approaches the XX model \cite{McCoy2} as $\gamma\rightarrow0$.
The XX model belongs to a different universality class and we will
therefore focus only on the parameter range $0<\gamma\leq1$.

The Hamiltonian (\ref{xymodel}) is symmetric under a global $\pi$
rotation about the $z$ axis ($\sigma^{x(y)}\rightarrow-\sigma^{x(y)}$),
which usually implies a zero value for the magnetization in the $x$
or $y$ direction ($\langle\sigma^{x(y)}\rangle=0$). However, as
the magnetic field $h$ is decreased (or $J$ increased) this symmetry
is spontaneously broken in the ground state (in the thermodynamic
limit) at $\lambda=J/h=\lambda_{1}\equiv1$, the first critical point
(1CP). More specifically, the ground state is doubly degenerate with
a finite magnetization ($\langle\sigma^{x}\rangle=\pm M$) in the
$x$ direction characterizing a ferromagnetic phase. It is also possible
to \emph{define} a symmetric ground state ($\langle\sigma^{x}\rangle=0$)
using a superposition of the two degenerate ones. Nonetheless, symmetric
\emph{macroscopic} states are just a theoretical construction with no
physical existence, since spontaneous symmetry-breaking mechanisms (superselection)
rapidly destroy such coherent superpositions (Schr\"{o}dinger cats)
in the thermodynamic limit \cite{Anderson}. These unphysical states
are called here symmetric states in contrast with the realistic broken-symmetry
ones ($\langle\sigma^{x}\rangle=\pm M$). Note that in the paramagnetic
phase ($\lambda\leq1$) there exists no such distinction.

By further decreasing the magnetic field a second phase transition
occurs at $\lambda=\lambda_{2}\left(\gamma\right)\equiv1/\sqrt{1-\gamma^{2}}$,
the second critical point (2CP). For magnetic fields smaller than
this critical value, the correlation functions do not tend to their
limiting value monotonically but in an oscillatory fashion \cite{McCoy2}.
The Ising limit, $\gamma=1$, exhibits only the first critical point.

As we will show in the following two Sections, for the calculation
of the bipartite and multipartite ($G(1)$ and $G(2,n)$) entanglement
all we need is the reduced density matrix of two spins, a $4\times4$
matrix that can be expanded in tensor products of Pauli matrices and
the identity $\sigma^{0}$: \begin{equation}
\rho_{i,j}=\frac{1}{4}\sum_{\alpha,\beta}p_{i,j}^{\alpha,\beta}\sigma_{i}^{\alpha}\otimes\sigma_{j}^{\beta},\end{equation}
 where \begin{equation}
p_{i,j}^{\alpha,\beta}=Tr[\sigma_{i}^{\alpha}\otimes\sigma_{j}^{\beta}\rho_{i,j}]=\langle\sigma_{i}^{\alpha}\otimes\sigma_{j}^{\beta}\rangle.\end{equation}
 The reduced density matrix $\rho_{i,j}$ is obtained by tracing out
all spins other than $i$ and $j$.

Remembering that $\rho_{i,j}$ is Hermitian with a unitary trace we
are left with nine independent matrix elements for $\rho_{i,j}$,
which are functions of the nine possible one and two-point correlation
functions ($p_{i,j}^{\alpha,\beta}=p_{i,j}^{\beta,\alpha}$). This
number can be further reduced by the symmetries of the problem. In
the XY model the global phase flip symmetry (global $\pi$ rotation
about the $z$ axis) in the paramagnetic phase ($\lambda\leq1$) implies $[\sigma_i^z\sigma_j^z,\rho_{i,j}]=0$, which imposes that $\langle\sigma_{i}^{x(y)}\rangle=\langle\sigma_{i}^{x}\sigma_{j}^{z}\rangle=\langle\sigma_{i}^{y}\sigma_{j}^{z}\rangle=0$, leaving only five independent
correlation functions: $\langle\sigma_{i}^{z}\rangle,\langle\sigma_{i}^{\alpha}\sigma_{j}^{\alpha}\rangle,\alpha=x,y,z$, and $\langle\sigma_{i}^{x}\sigma_{j}^{y}\rangle$.
In the ferromagnetic phase ($\lambda>1$) this no longer holds since
the Hamiltonian symmetry is not preserved by the ground state and
we have to explicitly evaluate the nine one and two-point correlation
functions.  $\langle\sigma_{i}^{z}\rangle,\langle\sigma_{i}^{\alpha}\sigma_{j}^{\alpha}\rangle,\alpha=x,y,z$ and $\langle\sigma_{i}^{x(y)}\rangle$
were obtained in \cite{McCoy1,McCoy2}. We are left then with three
off-diagonal two-point correlation functions to calculate, $\langle\sigma_{i}^{x}\sigma_{j}^{y}\rangle$,
$\langle\sigma_{i}^{x}\sigma_{j}^{z}\rangle$ and $\langle\sigma_{i}^{y}\sigma_{j}^{z}\rangle$.
Finally, due to the translational symmetry of the model, $\rho_{i,j}$
depends only on the distance $n=|i-j|$ between the spins, $p_{i,j}^{\alpha,\beta}=p_{n}^{\alpha,\beta}$
and $p_i^{\alpha,0}\equiv p_{i}^{\alpha}=p^{\alpha}$ is the same for all spins.

\begin{figure}
\begin{center}\includegraphics[%
  scale=0.8]{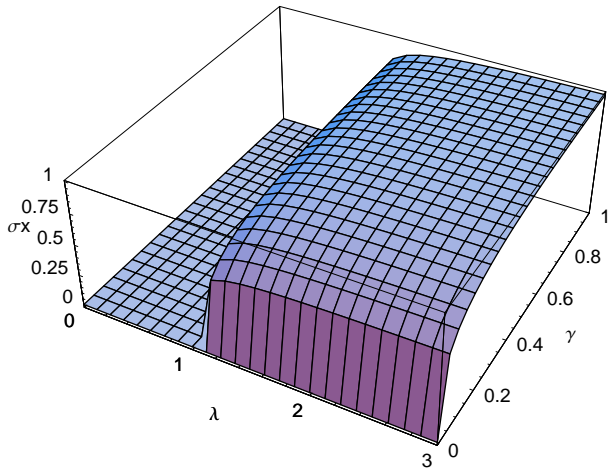}\end{center}

\caption{\label{fig1} (Color online) Magnetization along the $x$-axis for the XY model, with anisotropy. The first transition, 1CP,  is apparent from the discontinuity at $\lambda=1$, for any anisotropy $\gamma$.}
\end{figure}

\begin{figure}
\begin{center}\includegraphics[%
  scale=0.8]{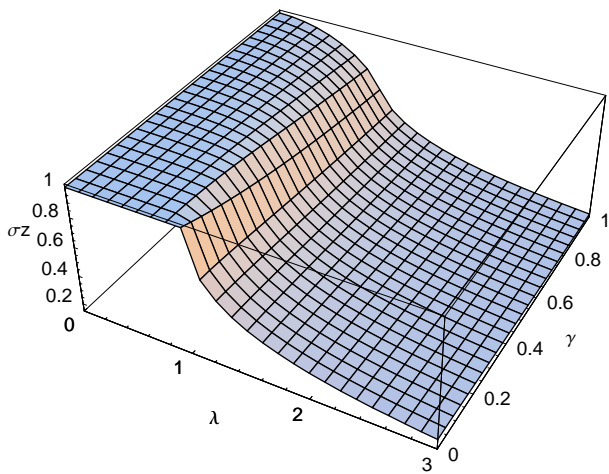}\end{center}

\caption{\label{fig2} (Color online) Magnetization along the $z$-axis for the XY model.}
\end{figure}

\begin{figure}
\begin{center}\includegraphics[%
  scale=0.8]{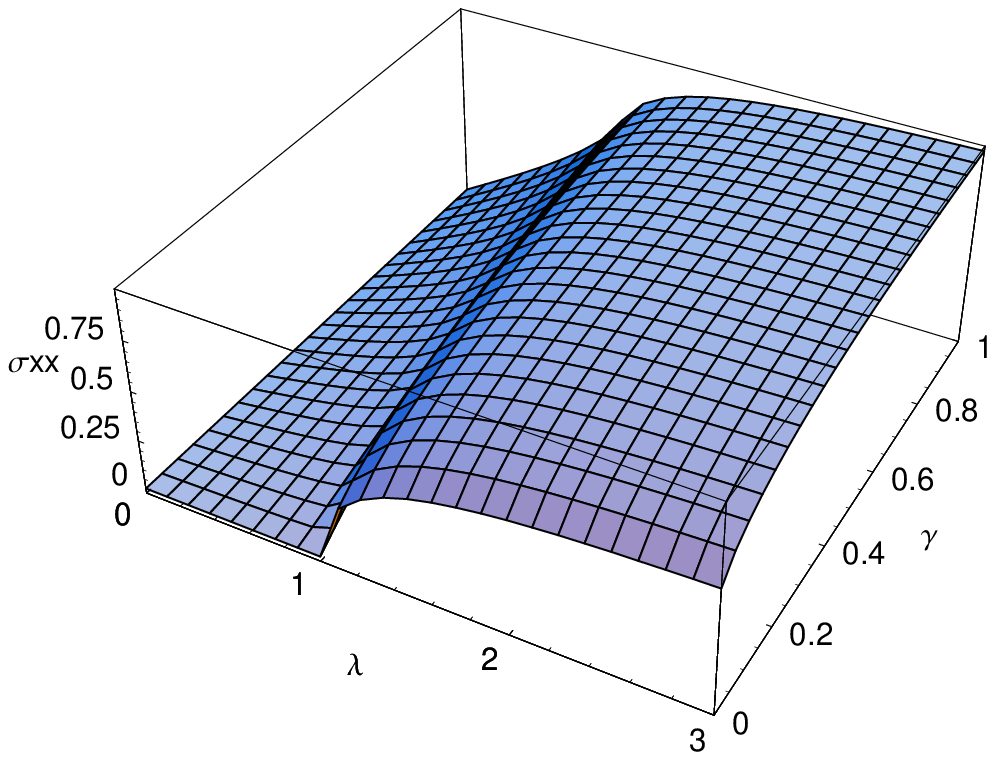}\end{center}

\caption{\label{fig3} (Color online) Nearest neighbor ($j=i\pm1$) diagonal correlation function $\langle\sigma_i^x\sigma_j^x\rangle$ for the XY model.}
\end{figure}

\begin{figure}
\begin{center}\includegraphics[%
  scale=0.8]{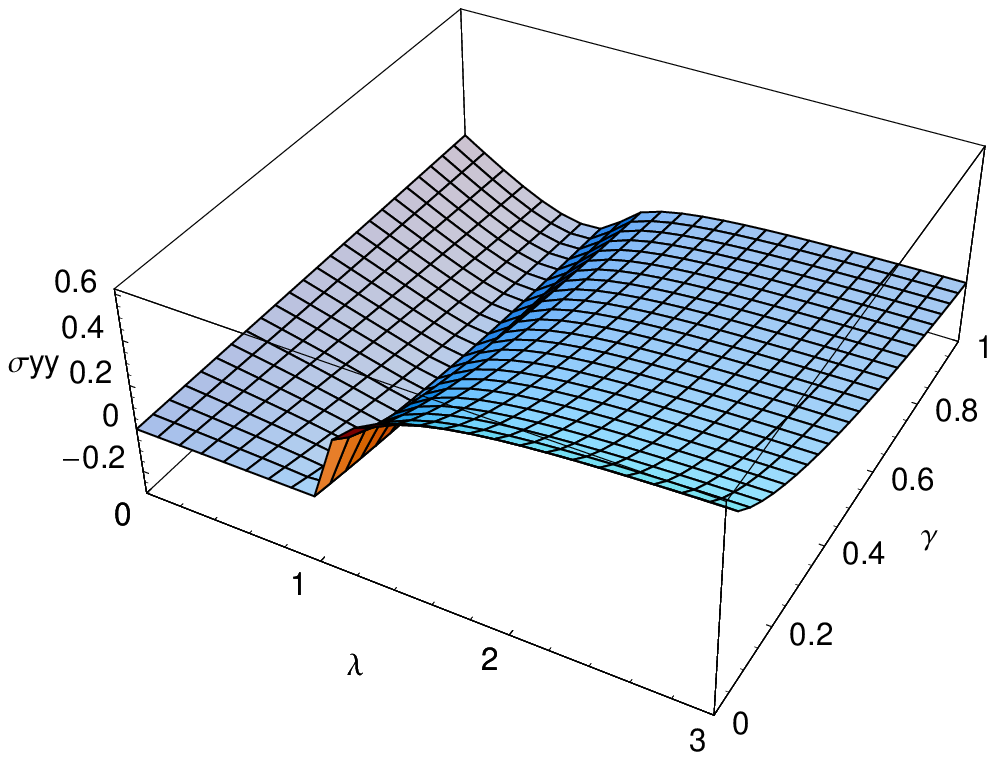}\end{center}

\caption{\label{fig4} (Color online) Nearest neighbor ($j=i\pm1$) diagonal correlation function $\langle\sigma_i^y\sigma_j^y\rangle$ for the XY model.}
\end{figure}

\begin{figure}
\begin{center}\includegraphics[%
  scale=0.8]{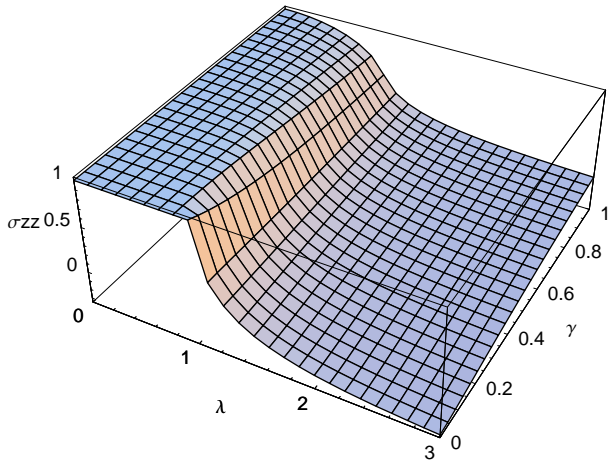}\end{center}

\caption{\label{fig5} (Color online) Nearest neighbor ($j=i\pm1$) diagonal correlation function $\langle\sigma_i^z\sigma_j^z\rangle$ for the XY model.}
\end{figure}

We could be tempted to say that $\rho_{i,j}$ is real since the matrix
elements of Hamiltonian~(\ref{xymodel}) are all real, and use this
fact to eliminate $p_{n}^{xy}$ and $p_{n}^{yz}$ as both quantities
appear in $\rho_{i,j}$ multiplied by the imaginary $i$. However, this
argument can be misleading since this {}``symmetry'' is not preserved
in the ferromagnetic state in the thermodynamic limit. As a
counterexample consider for instance the Ising Hamiltonian with the
nearest-neighbor coupling in the $y$ direction and a magnetic field in
the $z$ direction ($\gamma=-1$). In the ferromagnetic phase we would
have a finite value for $\langle\sigma_{i}^{y}\rangle$, which then
results in complex elements in $\rho_{i,j}$. Fortunately, exact
expressions for two of the three remaining off-diagonal correlation
functions have been obtained by Johnson and McCoy \cite{McCoy3}. In
fact, they have calculated the full time-dependent correlation
functions $\langle\sigma_{i}^{y}(0)\sigma_{i}^{z}(t)\rangle$ and
$\langle\sigma_{i}^{x}(0)\sigma_{i}^{z}(t)\rangle$. In particular,
they have shown that at any time $t$,
$\langle\sigma_{i}^{y}(0)\sigma_{i}^{z}(t)\rangle=0$ for all values of
$\gamma$ and $h$, which leads to $p_{n}^{yz}=0$. For
$\langle\sigma_{i}^{x}(0)\sigma_{i}^{y}(t)\rangle$, they have shown
that the leading term for large $n$ in both phases is linear in $t$,
suggesting that $p_{n}^{xy}$ ($t=0$) might be exactly zero. Numerical
calculations of $p_{n}^{xy}$ for small chains have confirmed that it
does indeed vanish in both phases \cite{andre}.  Collecting all the
previous results the reduced two-spin density matrix $\rho_{i,j}$ can
be written as
\begin{equation} \frac{1}{4}\!\!\left(\!\!\begin{array}{cccc}
      1+2p^{z}+p_{ij}^{zz} & p^{x}+p_{ij}^{xz} & p^{x}+p_{ij}^{xz} & p_{ij}^{xx}-p_{ij}^{yy}\\
      p^{x}+p_{ij}^{xz} & 1-p_{ij}^{zz} & p_{ij}^{xx}+p_{ij}^{yy} & p^{x}-p_{ij}^{xz}\\
      p^{x}+p_{ij}^{xz} & p_{ij}^{xx}+p_{ij}^{yy} & 1-p_{ij}^{zz} & p^{x}-p_{ij}^{xz}\\
      p_{ij}^{xx}-p_{ij}^{yy} & p^{x}-p_{ij}^{xz} & p^{x}-p_{ij}^{xz}
      & 1-2p^{z}+p_{ij}^{zz}\end{array}\!\!\right)\!.\end{equation}
The last off-diagonal correlation function, $p_{n}^{xz}$, was obtained
in terms of cumbersome complex integrals in \cite{McCoy3} rendering
its explicit computation very tedious. However, we were able to obtain
bounds for it from the physical restriction that all eigenvalues of
$\rho_{i,j}$ must be positive. Considering one of its eigenvalues as a
function of $p_{n}^{xy}$ results in a second-degree polynomial with
negative second derivative (we have checked this for many values of
$\lambda$ ranging from $0$ to $3$ and for $\gamma$ ranging from $0.1$
to $1$). This allowed us to obtain tight lower and upper bounds for
the value of $p_{n}^{xz}$.  This completes our construction of the
reduced density matrix of two spins, which is all we need for the
calculation of entanglement.  In Figs. (\ref{fig1})-(\ref{fig5}) we
plot the magnetization $\langle\sigma_{i}^{\alpha}\rangle$ along
$\alpha=x,z$ and the diagonal correlation functions,
$\langle\sigma_{i}^{\beta}\sigma_{j}^{\beta}\rangle,\,\beta=x,y,z$,
for nearest neighbors, $j=i\pm1$, for later discussion.

\section{Bipartite Entanglement}

\label{sec:Bipartite-Entanglement}

The bipartite entanglement between a pair of two-level systems (qubits
or $S=1/2$ spins) can be quantified using the concurrence $C$, since
it is a monotonic function of the entanglement of formation \cite{wootters},
a well established measure. The concurrence can be obtained from the
density matrix of the two spins and is given by $C=\text{max}\{0,\epsilon_{1}-\epsilon_{2}-\epsilon_{3}-\epsilon_{4}\}$,
where $\epsilon_{i}$, $i=1,\ldots,4$, are the square roots of the
eigenvalues, in decreasing order, of the matrix $R=\rho\tilde{\rho}$.
Here $\tilde{\rho}=(\sigma^{y}\otimes\sigma^{y})\rho^{*}(\sigma^{y}\otimes\sigma^{y})$.
Another pairwise measure of entanglement is the negativity which is
based on the Peres-Horodecki separability test \cite{peres,horodecki}.
This test states that a separable state is always positive under partial
transposition (PPT). This is also a sufficient condition for separability
in the case of two-level systems. Thus, it is reasonable to quantify
entanglement measuring {}``how much'' the partially transposed density
matrix is negative. A possible definition of negativity, which was
proved to be an entanglement monotone \cite{werner}, is given as
$N(n)=\text{max}\{0,-2\,\text{min}(u_{k})\}$, where $u_{k}$ are
the eigenvalues of the partial transpose of $\rho_{i,i+n}$ and the
label $n$ denotes the distance between the qubits. The main advantage
of the negativity over the concurrence is that the former is easier
to compute than the latter.

In the paramagnetic phase ($\lambda\leq1$) or for any value of $\lambda$
in the symmetric ground state (i. e., without symmetry breaking), the
concurrence and the negativity are simple expressions in terms of
the correlation functions. Mathematically, this is a consequence of
the fact that the fourth-degree equations resulting from the diagonalization
of $R$ factorize into two equations of the second degree \cite{dinamarques}.
The expression for the concurrence, valid for any system possessing
the same symmetries as the symmetric ground state of the XY model
is written as\begin{equation}
C(n)=\text{max}\{0,C'(n),C''(n)\},\label{concurrence}\end{equation}
 where\begin{eqnarray}
C'(n) & = & \frac{1}{2}(|p_{n}^{xx}-p_{n}^{yy}|+p_{n}^{zz}-1),\\
C''(n) & = & \frac{1}{2}\left(|p_{n}^{xx}+p_{n}^{yy}|-\sqrt{(1+p_{n}^{zz})^{2}-4(p^{z})^{2}}\right).\end{eqnarray}
 The negativity expression derived exclusively for the XY model reads
\begin{equation}
N(n)=\text{max}\{0,-2\,\text{min}[u_{1}(n),u_{3}(n)]\},\label{negativity}\end{equation}
 where \begin{eqnarray}
u_{1}(n) & = & -\frac{1}{2}\left(1+p_{n}^{zz}-\sqrt{(p_{n}^{xx}+p_{n}^{yy})^{2}+4(p^{z})^{2}}\right),\\
u_{3}(n) & = & -\frac{1}{2}(1-p_{n}^{xx}+p_{n}^{yy}-p_{n}^{zz}).\end{eqnarray}
 The expression for the concurrence could be written in this general
form (independent of the particular values of the one and two-point
correlation functions) because it is derived simply by imposing the
positivity of $\rho_{i,j}$ and the fact that all eigenvalues of $R$
are real numbers. We have also observed that for $\gamma^{2}+1/\lambda^{2}>1$, i.e.
$\lambda<\lambda_{2}\left(\gamma\right)$ it is always
$C'(n)$ and $u_{1}(n)$ that are relevant for the concurrence and
the negativity through Eqs.~(\ref{concurrence}) and (\ref{negativity}),
respectively. On the other hand, for $\gamma^{2}+1/\lambda^{2}<1$, i.e.
$\lambda>\lambda_{2}\left(\gamma\right)$, it is $C''(n)$
and $u_{3}(n)$ that appear in these two measures, respectively. The
change from $\gamma^{2}+1/\lambda^{2}>1$ to $\gamma^{2}+1/\lambda^{2}<1$
occurs at the 2CP, i.e. at $\lambda=\lambda_{2}\left(\gamma\right)=1/\sqrt{1-\gamma^{2}}$.

In the ferromagnetic phase (for the broken-symmetry state) the calculation
of the concurrence is not so simple since we have a fourth-degree
equation which factorizes into a first-degree one and a complicated
third-degree equation. Although the latter can be solved exactly,
the expressions for its roots are not very illuminating, rendering
a detailed general analysis unfeasible. Fortunately, it was demonstrated
\cite{dinamarques} that for the Ising model the concurrence does
not change upon spontaneous symmetry breaking. This opened the possibility
for the use of the simple expression of the paramagnetic phase in
the ferromagnetic one. The analysis can be extended to the XY model
since the reduced density matrices of the two models have a similar form.
The condition for an identical expression for the concurrence in the
paramagnetic and ferromagnetic phases is \begin{equation}
\sqrt{(1+p_{n}^{zz})^{2}-4(p^{z})^{2}}+p_{n}^{zz}-2p_{n}^{yy}-1>0.\label{condition}\end{equation}

\begin{figure}
\begin{center}\includegraphics[%
  scale=0.8]{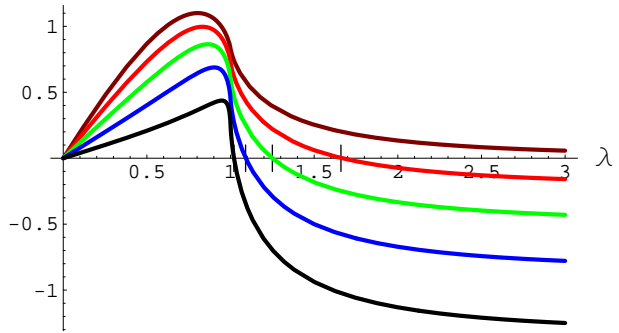}\end{center}

\caption{\label{fig6} (Color online) Plot of the left-hand side of Eq.~(\ref{condition})
as a function of $\lambda$: when the function is positive the concurrence
for the XY model does not change upon symmetry breaking. The anisotropies
are $\gamma=1$, $0.8$, $0.6$, $0.4$, and $0.2$ (from top to bottom).
We have also plotted a small vertical bar to represent the position
of the 2CP ($\lambda_{2}\left(\gamma\right)=1/\sqrt{1-\gamma^{2}}$).}
\end{figure}

In Fig.~\ref{fig6} we show the left-hand side of Eq.~(\ref{condition})
as a function of $\lambda$ for the XY model. It can be seen that
Eq.~(\ref{condition}) always holds for the Ising model ($\gamma=1$,
first curve from top to bottom) but is violated after the 2CP ($\lambda>\lambda_{2}\left(\gamma\right)=1/\sqrt{1-\gamma^{2}}$)
in the XY model ($\gamma\neq1$). The critical value $\lambda_{2}\left(\gamma\right)$
has been indicated by the vertical lines for each $\gamma$ in Fig.~\ref{fig6}.

\begin{figure}
\begin{center}\includegraphics[%
  scale=0.8]{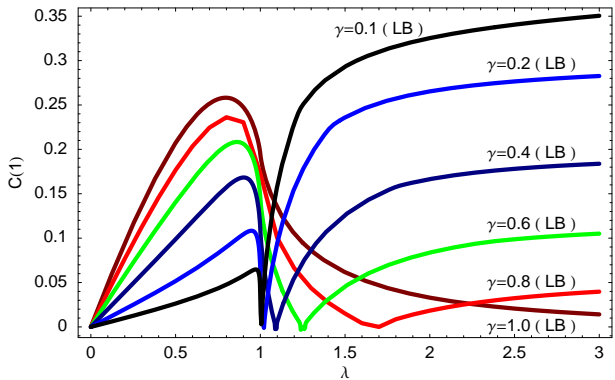}\end{center}

\caption{\label{fig7} (Color online) Lower bound of the concurrence for nearest
neighbors obtained using the upper bound of $p_{1}^{xz}$. In the
limit of small magnetic field (large $\lambda$) the entanglement
decreases with increasing $\gamma$. Here $\gamma=1,0.8,0.6,0.4,0.2$,
and $0.1$ from top to bottom in the $\lambda\leq1$ phase.}
\end{figure}

\begin{figure}
\begin{center}\includegraphics[%
  scale=0.8]{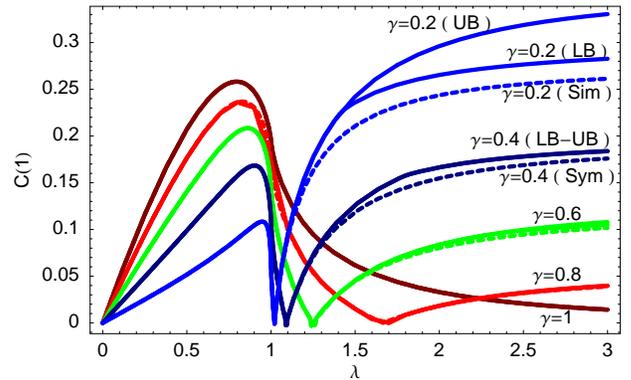}\end{center}

\caption{\label{fig8} (Color online) The lower and upper bounds for the nearest-neighbor
concurrence (solid lines) and the concurrence in the symmetric state
(dashed lines). Here $\gamma=1,0.8,0.6,0.4$, and $0.2$ from top
to bottom in the $\lambda\leq1$ phase. Note that for most of the
anisotropies we can barely see the difference between the lower and
upper bounds.}
\end{figure}

Using the above expressions and the correlation functions depicted in
Figs. (\ref{fig1})-(\ref{fig5}), including the bounds for $p_{n}^{xz}$
obtained through the procedure explained in
Section~\ref{sec:XY-model}, we now analyze the bipartite entanglement
between any two spins of the XY chain. Notice that for the following
discussion we have calculated numerically the concurrence, not relying
in the simplified formula (\ref{concurrence}). First, we have
evaluated the concurrence of nearest neighbors ($C(1)$) for some
values of the anisotropy $\gamma$. These results are shown in
Fig.~\ref{fig7}, where we plot the lower bounds for the concurrence.
Moreover, in Fig.~\ref{fig8} we show both the lower and the upper
bounds for the broken-symmetry state as well as the concurrence in the
unphysical symmetric ground state. As it is already known
\cite{nielsen,Nature}, $C(1)$ is not maximal at the 1CP. It is
important to note that the bounds are very tight near the QPTs
allowing us to correctly characterize the behavior of the concurrence
at the CPs. Indeed, Fig.~\ref{fig8} shows that the concurrence changes
more abruptly (a diverging derivative) at the 2CP than at the 1CP
where the spontaneous symmetry breaking occurs
($\lambda=\lambda_{1}=1$). We can also see that after the 1CP the
concurrence starts to decrease, vanishing at the 2CP. This fact had
already been observed \cite{tognetti} and it can be shown that at this
point the ground state is completely separable \cite{tognetti}.

\begin{figure}
\begin{center}\includegraphics[%
  scale=0.8]{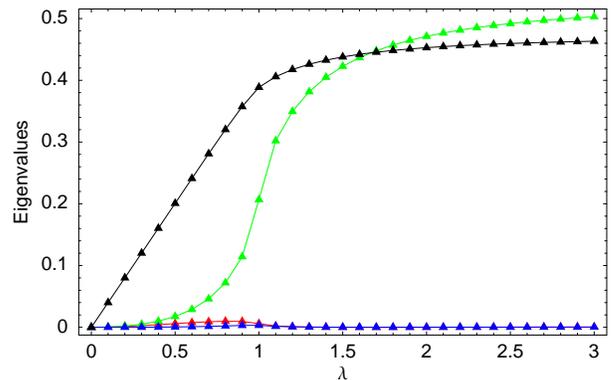}\end{center}

\caption{\label{fig9} (Color online) Plot of the square root of the eingenvalues
of the matrix R (for $\gamma=0.8$), used to obtain the concurrence,
for the symmetric case. We can observe a change in the greatest eigenvalue
of R at the second critical point resulting in a change from $C'(n)$
to $C''(n)$ as the expression that contributes to the concurrence.}
\end{figure}

\begin{figure}
\begin{center}\includegraphics[%
  scale=0.8]{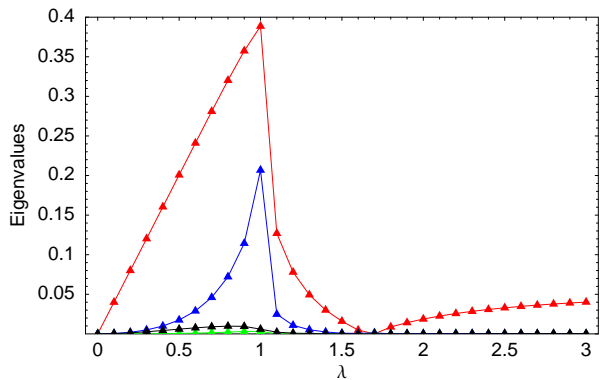}\end{center}

\caption{\label{fig10} (Color online) Plot of the square root of the eingenvalues
of the matrix R (for $\gamma=0.8$), used to obtain the concurrence,
for the broken-symmetry case. We can observe that for the broken-symmetry
case there is no crossing of eigenvalues of R at the second critical
point. Actually, all eigenvalues vanish at the 2CP. This plot was
obtained using the upper bound for $p^{xz}$ but there is no visible
difference if we use the lower bound.}
\end{figure}

Remarkably, the discrepancy between the symmetric and the
broken-symmetry cases, in contrast to what one might expect, only
occurs after the 2CP ($\lambda>\lambda_{2}$), where the correlation
functions tend to their limiting value in an oscillatory fashion
\cite{McCoy2}. Thus, the spontaneous symmetry breaking, which occurs
already at the 1CP, has no influence on bipartite entanglement.  Even
after the 2CP the difference between the symmetric and the
broken-symmetry states is small and becomes more pronounced only as
$\gamma\rightarrow0$, where the former has slightly \emph{less}
entanglement than the latter. Finally, the {}``origin'' of the
entanglement is different in the two states since in the symmetric
case at the 2CP there is a change in the greatest eigenvalue of $R$,
see Fig.~\ref{fig9} (this results in a change from $C'(n)$ to $C''(n)$
as the expression that contributes to the concurrence). This has been
interpreted \cite{tognetti} as a change in the kind of entanglement
present in the ground state. However this is not true when one
correctly employs the broken-symmetry state since now it is the same
eigenvalue of $R$ that is maximal for all values of $\lambda$, see
Fig. ~\ref{fig10}. We can also see that, as we approach the XX model
($\gamma\rightarrow0$), the concurrence decreases in the $\lambda<1$
region, vanishing at the 1CP, and then increases in the ferromagnetic
phase ($\lambda>1$).

Similar conclusions can be derived for the negativity since the curves
for the negativity as a function of $\lambda$ and $\gamma$ are very
close to the ones already shown for the concurrence. Here we only
plot the negativity for nearest neighbors, $N(1)$, in the symmetric
state in Fig.~\ref{fig6}. We should note that, as for the concurrence,
the negativity of the symmetric ground state is close to the negativity
of the broken-symmetry state in the ferromagnetic phase.
\begin{figure}\begin{center}\includegraphics[scale=0.8]{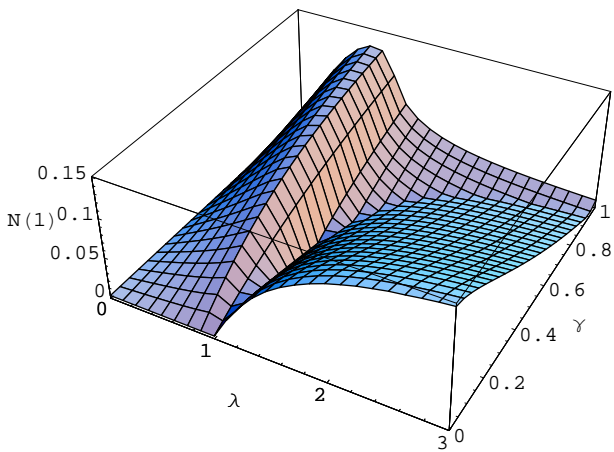}\end{center}
\caption{\label{fig11} (Color online) Nearest neighbor negativity in the symmetric
ground state. We can see that at the critical point the entanglement
decreases as we approach the XX model ($\gamma\rightarrow0$).}
\end{figure}

\begin{figure}
\begin{center}\includegraphics[%
  scale=0.8]{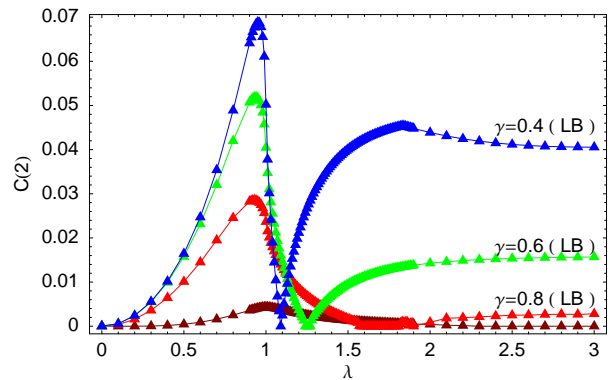}\end{center}

\caption{\label{fig12} (Color online) Lower bound for the concurrence for
next-nearest neighbors using the upper bounds of the off-diagonal
correlation function $p_{2}^{xz}$. Here $\gamma=1,0.8,0.6$, and
$0.4$ from bottom to top in the $\lambda\leq1$ phase. The broken-symmetry
state was used in the ferromagnetic phase ($\lambda>1$).}
\end{figure}

\begin{figure}
\begin{center}\includegraphics[%
  scale=0.8]{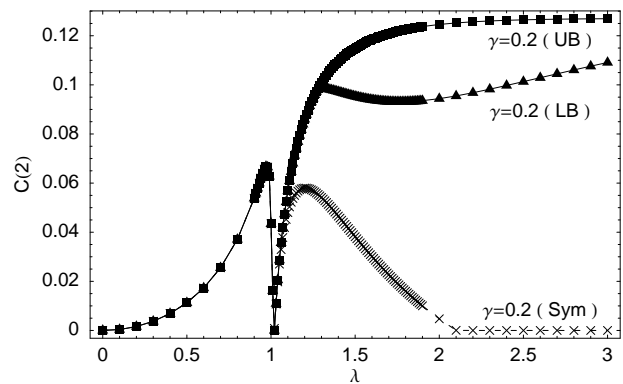}\end{center}

\caption{\label{fig13} (Color online) Concurrence for next-nearest neighbors
in the broken-symmetry state at $\gamma=0.2$, obtained using the
upper (triangles) and lower (squares) bounds of the off-diagonal correlation
function $p_{2}^{xz}$ . For comparison, we also plot the concurrence
in the symmetric state (crosses).}
\end{figure}

We have also obtained the concurrence between next-nearest neighbors
$C(2)$ in the broken-symmetry state. In Fig.~\ref{fig12} we show
its lower bound for some values of $\gamma$. We can see that in contrast
to $C(1)$, $C(2)$ at first increases in the paramagnetic phase as
we leave the Ising model in the direction of the XX model. As a function
of $\lambda$, $C(2)$ reaches its maximal value just before $\lambda=1$
(the 1CP), which increases as $\gamma\rightarrow0$. For small magnetic
field (large $\lambda$) we see the same behavior as for $C(1)$:
the entanglement increases as we approach the XX model ($\gamma\rightarrow0$).
The difference in the entanglement of the symmetric and the broken-symmetry
states, however, is more pronounced now. In contrast to the broken-symmetry
state, the entanglement in the symmetric state vanishes for $\lambda$
larger than a certain value. This is illustrated in Fig.~\ref{fig13}
where we show the lower and upper bounds for the concurrence in the
broken-symmetry state compared to the concurrence in the symmetric
state for $\gamma=0.2$. We should note that for all values of $\gamma>0.2$,
the bounds are tighter in comparison.

\begin{figure}
\begin{center}\includegraphics[%
  scale=0.8]{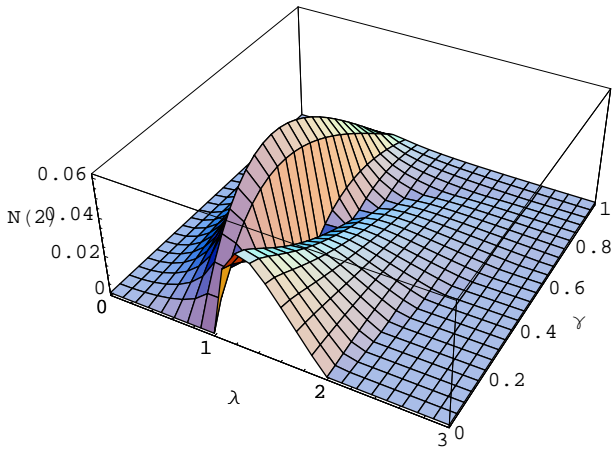}\end{center}

\caption{\label{fig14} (Color online) Negativity for next-nearest neighbors
in the symmetric ground state. We can see that at the first critical
point $\left(\lambda=1\right)$ the entanglement first increases as
we leave the XX model ($\gamma=0$) and then starts to decrease for
sufficiently large $\gamma$.}
\end{figure}

The negativity for next-nearest neighbors has a similar behavior but
with smaller values in the region of small fields (large $\lambda$).
This can be better viewed in Fig.~\ref{fig14} where we plot $N(2)$
in the symmetric state. Note that in the symmetric state the negativity
(and also the concurrence) vanishes for sufficiently small values
of the field.

We have also calculated the concurrence and the negativity for spins
three and four lattice sites apart, which show behaviors similar to
$C(2)$ and $N(2)$. The only difference is that the entanglement
is much smaller, decreasing as we increase the distance between the
spins. It should be noted that the concurrence and negativity values
for next-nearest neighbors are significantly smaller than the ones
for nearest neighbors, which means that bipartite entanglement is
more concentrated on nearest-neighbor sites.

One last fact about bipartite entanglement which we would like
to mention is related to the origin of the non-analyticities of the
concurrence and of the ground state energy per site $\mathcal{E}$.
It terms of the two spin reduced density matrix elements
$(\rho_{i,j})_{\alpha \beta}=(\rho)_{\alpha\beta}$ we have
\begin{equation}
\partial^2_{\lambda}\mathcal{E} = -\frac{2}{\lambda}\partial_{\lambda}
[(\rho)_{22}+(\rho)_{44}]
\end{equation}
and
\begin{equation}
C(1)=2[(\rho)_{41}-(\rho)_{22}],
\end{equation}
where C(1) is the concurrence for the symmetric case and $\partial^2_{\lambda}$
stands for the second order derivative with respect to the tuning
parameter $\lambda$. 
As argued in
\cite{sarandy} both $\partial_{\lambda}^2 \mathcal{E}$ and
$\partial_{\lambda} C(1)$ exhibit critical behavior through
their dependece upon $\partial_{\lambda}(\rho)_{22}$, since
$(\rho)_{41}$ is well behaved. Looking at the
correlation functions we observe that the divergence of
$\partial_{\lambda}(\rho)_{22}$ comes from the correlation 
in the $z$ direction as $(\rho)_{22}=1-p^{zz}$. However
$(\rho)_{44}$ also has a dependence on $p^{zz}$ which cancels
the one in $(\rho)_{22}$, and we end up with
\begin{equation}
\partial_{\lambda}^2 \mathcal{E}=-\frac{1}{\lambda}\partial_{\lambda}
p^z.
\end{equation}
Therefore, in terms of the correlation functions, the critical behavior of
$\partial_{\lambda}^2 \mathcal{E}$
is originated on $p^z$ while the divergence of $\partial_{\lambda} C(1)$
is given by $p^{zz}$, since $(\rho)_{41}=p^{xx}-p^{yy}$ does
not cancel the divergence of $p^{zz}$.

In sum, based on these measures of entanglement we can say that there
is more bipartite entanglement around the 1CP when we approach the
Ising model. When we approach the XX model the entanglement is more
pronounced in the ferromagnetic phase. We expect this behavior to
continue to hold for $C(n)$ and $N(n)$ when $n>4$. It has already been
pointed out that the range in which the concurrence has a finite value
increases as 1/$\gamma$ \cite{Nature}, being infinite for the XX
model. Finally, we would like to stress that the concurrence
does not change upon symmetry breaking in the Ising model. 
This fact shows that it suffers no influence during the symmetry-breaking
process, although bipartite entanglement is able to mark the phase
transition through a diverging derivative of the concurrence.

In the XY model, on the other hand, the concurrence
is different for
symmetric and broken-symmetry states. However, this difference only
appears at the second critical point, not as a result of the symmetry
breaking that occurs at the first critical point. Thus, we have shown
that for both the Ising and the XY model, the symmetry-breaking phase
transition does not affect the bipartite entanglement. In fact, as
three of us have already argued \cite{newmeasure}, the fact that the
concurrence does not depend on the magnetization in the $x$ direction
$p^{x}$ is a possible explanation for the fact that it is not maximal
at the critical point. 

\section{Multipartite Entanglement}

\label{sec:Multipartite-Entanglement} 

After studying the entanglement between two spins in the chain we
now focus our attention on the multipartite entanglement (ME) in the
XY model as given by $G(1)$ \cite{meyer} and $G(2,n)$ \cite{nosso paper 1,newmeasure,nosso paper 3}. For a spin-1/2 chain these are given by \begin{equation}
G(1)=\frac{d}{d-1}\left[1-\frac{1}{N}\sum_{j=1}^{N}\text{Tr}\left(\rho_{j}^{2}\right)\right]\label{g1}\end{equation}
 and \begin{equation}
G(2,n)=\frac{d}{d-1}\left[1-\frac{1}{N-n}\sum_{j=1}^{N-n}\text{Tr}\left(\rho_{j,j+n}^{2}\right)\right],\label{g2}\end{equation}
 where $d$ is the dimension
of the Hilbert space of the reduced density matrices $\rho_{j}$ or
$\rho_{j,j+n}$, i.e., $d=2$ for $G(1)$ and $d=4$ for $G(2,n)$.
For systems with translational invariance
such as the XY model, these measures simply reduce to the the linear entropy of one spin and two spin $n$-sites apart, respectively.

One advantage of the measures (\ref{g1}) and (\ref{g2}) is that they
are simple to evaluate since we just need the reduced one- and
two-spin density matrices that were already obtained previously. This
kind of measure has received the name of local entanglement
measure/estimator since they depend on the reduced density operator of
two spins, which is a local quantity. Moreover, inspired by accumulated
experience of many-body physics, we believe that a great deal can be
learned from the knowledge of two-point correlation functions only.
$G(2,n)$ inherits the
full non-analytical behavior of the elements of the reduced density
matrix of two spins and it is possible to make a general link between
divergences in the derivatives of the energy, which signal QPTs, and
$G(2,n)$ or its derivatives \cite{nosso paper 3}. In other words,
$G(2,n)$ and/or its derivatives are able to signal QPTs as do
bipartite entanglement measures, with the exceptions of the cases
where the non-analyticities accidentally cancel out. In
Ref.~\cite{nosso paper 3} it was shown that $G(1)$ and $G(2,n)$ are
maximal at the 1CP and zero at the 2CP, thus possessing the ability to
map out the complete phase diagram of the XY model. Here we intend to
analyze those results in vision of the differences between the
symmetric and broken-symmetry cases/states.

For the XY model Eqs. (\ref{g1}) and (\ref{g2}) above can be rewritten in terms of
the one and two-point correlation functions, depicted in Figs. (\ref{fig1})-(\ref{fig5}), as \begin{equation}
G(1)=1-\langle\sigma_{j}^{x}\rangle^{2}-\langle\sigma_{j}^{z}\rangle^{2}\label{g1b}\end{equation}
 and \begin{eqnarray}
G(2,n) & = & 1-\frac{1}{3}\left(2\langle\sigma_{j}^{x}\rangle^{2}+2\langle\sigma_{j}^{z}\rangle^{2}+2\langle\sigma_{j}^{x}\sigma_{j+n}^{z}\rangle^{2}+\right.\nonumber \\
 &  & \langle\sigma_{j}^{x}\sigma_{j+n}^{x}\rangle^{2}+\left.\langle\sigma_{j}^{y}\sigma_{j+n}^{y}\rangle^{2}+\langle\sigma_{j}^{z}\sigma_{j+n}^{z}\rangle^{2}\right).\label{g2b}\end{eqnarray}
 As we mentioned before, Eqs.~(\ref{g1b}) and (\ref{g2b}) have
already been shown to be maximal at the 1CP of the Ising \cite{nosso paper 1}
and XY \cite{nosso paper 3} models. We have argued that this behavior
is due to the emergence of a finite value of the magnetization with
the spontaneous symmetry breaking for $\lambda>1$, since this is
the correlation function in the expressions for $G(1)$ and $G(2,n)$
exhibiting the most abrupt change, despite being continuous at the
1CP. This will be made clearer as we investigate the behavior in the
symmetric case. We also note that at the 1CP $G(2,n)$ always increases
as a function of $n$ which is a strong indication of genuine ME.
The von Neumann entropy (Entropy of Entanglement) of one spin has
already been shown to be maximal at the 1CP \cite{nielsen}.

\begin{figure}
\begin{center}\includegraphics[%
  scale=0.8]{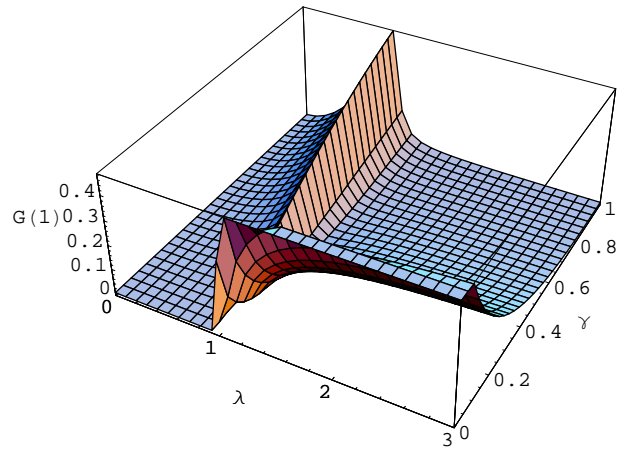}\end{center}

\caption{\label{fig15} (Color online) G(1) for the XY model in the broken-symmetry
case.}
\end{figure}

We first show $G(1)$ for the XY model as a function of $\lambda$
and $\gamma$ for the broken-symmetry case (see Fig.~\ref{fig15}).
We can see that for all values of the anisotropy $G(1)$ is maximal
at the 1CP and decreases as one approaches the XX model. We have also
checked that $G(1)$ is zero at the 2CP $\lambda=\lambda_{2}\left(\gamma\right)=1/\sqrt{1-\gamma^{2}}$ as expected, since at this
point the state is completely separable \cite{tognetti}. 

\begin{figure}
\begin{center}\includegraphics[%
  scale=0.8]{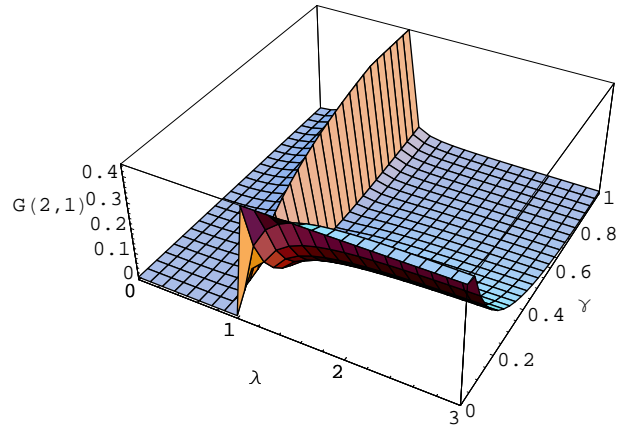}\end{center}

\caption{\label{fig16} (Color online) G(2,1) for the XY model in the broken-symmetry
case.}
\end{figure}

\begin{figure}
\begin{center}\includegraphics[%
  scale=0.8]{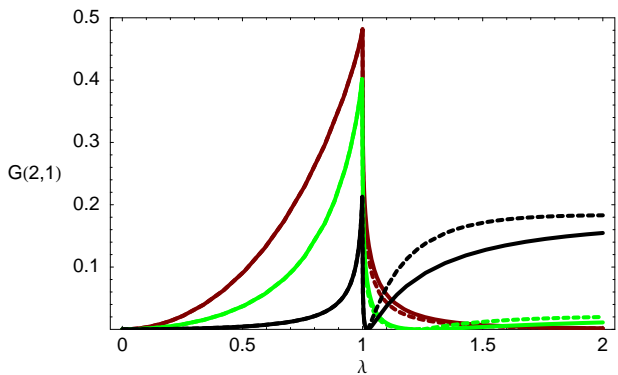}\end{center}

\caption{\label{fig17} (Color online) Lower (solid lines) and upper (dashed
lines) bounds for G(2,1) of the XY model in the broken-symmetry case
for three values of the anisotropy: $\gamma=1,0.6,0.2$. (from top
to bottom in the $\lambda<1$ region).}
\end{figure}

The upper bound for $G(2,1)$ for the broken-symmetry case is shown
in Fig.~\ref{fig16}. We note that it behaves similarly to $G(1)$.
To check the quality of the bounds we also plot the lower and upper
bounds for $G(2,1)$ in the broken-symmetry case for three values
of anisotropy in Fig.~\ref{fig17}. The same behavior was also found
for $G(2,2)$ and $G(2,7)$ \cite{nosso paper 3}. In fact, the value of $G(2,n)$ for a fixed $\lambda$ always increases as
a function of $n$, which is in contrast to bipartite entanglement, and also in contrast to the correlation functions, which decrease as $n$
increases. Therefore, $G(2,n)$ must increase (see Eq.~\ref{g2b}),
either as a power law at the critical point or exponentially away
from it. As mentioned before, this feature allows the definition of
an entanglement length (see Ref.~\cite{nosso paper 3} for more details)
and indicates that at the CP the entanglement is more distributed
in the chain than anywhere else.

Since both $G(1)$ and $G(2,n)$ essentially show the same behavior,
from Eq. (\ref{g1b}) we conclude that the magnetizations
$\langle\sigma_j^x\rangle$ and $\langle\sigma_j^z\rangle$ are the
minimal quantities from which the multipartite entanglement over the
chain can be inferred.  Now we observe that near the XX model
($\gamma\rightarrow 0$ ) the ferromagnetic phase shows more
entanglement than the CP, a feature observed for bipartite
entanglement between two spins as well, as given by the concurrence
and negativity (see Figs. (\ref{fig7}) and (\ref{fig11})). This
behavior should be contrasted with Figs. (\ref{fig1})-(\ref{fig5}) for
the magnetization and diagonal correlations. Notice that those one and
two-point correlations (apart from
$\langle\sigma_i^y\sigma_j^y\rangle$) are invariably smaller closer to
the XX model than closer to the Ising model, in particular
$\langle\sigma_j^x\rangle$ and $\langle\sigma_j^z\rangle$, showing
the well know fact that, deep in the ferromagnetic phase the magnetization
is destroyed by quantum fluctuations as ones goes from the
Ising Hamiltonian to the XX limit. Thus it is reassuring to see that
the proposed indicators of "quantum character", both bipartite and multipartite
entanglement measures, do indeed increase as we approach the XX model
from the Ising one. This
feature highlights the differences between classical and quantum
correlations (entanglement) in the XY model. Thus, although the
correlation functions involve both classical and quantum correlations,
only \emph{a proper combination} of them can reveal their entanglement
content.

\begin{figure}
\begin{center}\includegraphics[%
  scale=0.8]{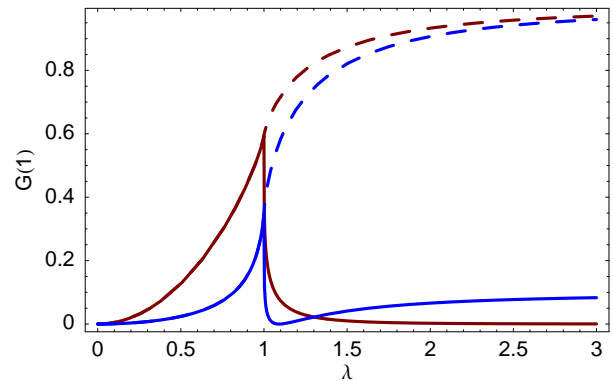}\end{center}

\caption{\label{fig18} (Color online) Comparison of
G(1) in the symmetric (dashed lines) and broken-symmetry (solid lines)
states for the XY model and for two values of anisotropy: $\gamma=1$ (brown)
and $0.4$ (blue).}
\end{figure}

\begin{figure}
\begin{center}\includegraphics[%
  scale=0.8]{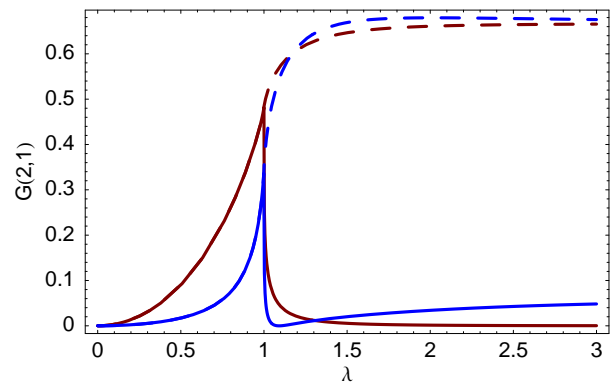}\end{center}

\caption{\label{fig19} (Color online) Comparison of G(2,1) in the symmetric
(dashed lines) and broken-symmetry (solid lines) states for the XY
model and for two values of anisotropy: $\gamma=1$ (brown) and $0.4$ (blue).}
\end{figure}

We now compare the symmetric and broken-symmetry states. In
Fig.~\ref{fig18} we have plotted $G(1)$ and in Fig.~\ref{fig19}
$G(2,1)$, both for two values of the anisotropy, $\gamma=1$ and
$\gamma=0.4$. It can be seen that in both cases the symmetric state
(dashed line) does not show a maximum at the 1CP. Instead, it is an
increasing monotonic function of $\lambda$. The crucial element here
is that the magnetization in the $x$ direction,
$\langle\sigma^{x}\rangle$, and
$\langle\sigma_{i}^{x}\sigma_{j}^{z}\rangle$ vanish in the region
$\lambda>1$ in the symmetric case, and the former is responsible for
the difference in behavior between $G(1)$ and $G(2,1)$. To check this,
we have made $\langle\sigma^{x}\rangle$ zero by hand in the expression
for $G(2,1)$ in the broken-symmetry case and verified that the result
is very similar to the symmetric case. This indicates that the
magnetization is the primary reason why $G(1)$ and $G(2,1)$ are
maximal at the 1CP and the results for the symmetric and
broken-symmetry cases are different. The reader should remember that
in the case of bipartite entanglement, given by the concurrence or the
negativity, spontaneous symmetry breaking had no effect, only
appearing to contribute in a different manner after the 2CP.
Furthermore, we observe that both $G(1)$ and $G(2,n)$ signal the two
critical points in the broken-symmetry state, while the same is not
true in the symmetric state. The complete phase diagram could thus be
drawn only by considering the non-analyticities of either of these two
measures. Thus symmetry-breaking has more effect over multipartite
entanglement. Again, the presence of the magnetization
$\langle\sigma^x\rangle$ (which is highly sensitive to
symmetry-breaking) in both $G(1)$ and $G(2,n)$ and its absence in the
concurrence and negativity is responsible for the different behavior
when one uses a broken symmetry ground state or not.

\section{Conclusions}

\label{sec:Conclusions}

We have extensively studied the entanglement properties of the one
dimensional XY model in a transverse magnetic field. We have in all
cases assumed the chain to be at $T=0$ and worked in the thermodynamic
limit of an infinite chain. One of our goals was to characterize both
the pairwise and multipartite entanglement of the ground state of
the XY model. In order to do a complete analysis we were forced to
consider two distinct ground states. The first one, which always preserves
all the symmetries of the Hamiltonian, was called the symmetric ground
state. This state, however, is unphysical for $\lambda=J/h>1$ since
in a realistic situation the global phase flip (global $\pi$ rotation
around the $z$ axis) is always spontaneously broken. Therefore, we
also considered a second state, namely the broken-symmetry ground
state, where this symmetry no longer holds.

For the broken-symmetry state we were able to show that, in contrast
to pairwise entanglement, multipartite entanglement is maximal at the
first critical point (where the XY model exhibits a diverging
correlation length). This property is not observed in the symmetric
state, in which case the multipartite entanglement increases
monotonically as we decrease the external magnetic field.
Furthermore, we have also shown that the concurrence does not change
in the vicinity of the symmetry-breaking critical point whether we
employ the symmetric or the broken-symmetry state. On the other hand,
we have explicitly shown that the multipartite entanglement depends
strongly on the symmetry of the ground state. This result suggests
that, as is the case for the XY model, the behavior of the
multipartite entanglement may be intimately connected with the
spontaneous symmetry breaking mechanism. We should also remark that
only after the second critical point is the concurrence dependent on
which state we use, and this is probably because the symmetry-breaking
has a more pronounced effect in the oscillatory behavior the
correlations show after this point.

We have arrived at another interesting result by noticing two
important facts. First, for spins three or more sites apart there
exists no pairwise entanglement whatsoever \cite{nielsen,Nature}.
Second, we have shown that multipartite entanglement is never zero at
the first critical point (being maximal for the broken-symmetry
state). Combining these two facts we are led to conclude that the long
range correlations at this critical point are a consequence of the
existence of multipartite entanglement and not pairwise entanglement.
Moreover both multipartite and bipartite entanglement tend to be
larger than at the 1CP in the ferromagnetic phase as one approaches
the XX model ($\gamma\rightarrow0$), increasing monotonically after
the 2CP. This feature contrasts with the one and two-point
correlations that tend to decrease as $\gamma\rightarrow0$, showing
that the enhanced quantum correlations (entanglement) in this region
can only be appreciated through a proper combination of the
correlation functions.  Since the XX model belongs to a different
universality class, nothing can be said about entanglement at
$\gamma=0$ from our study. It would be certainly interesting to
investigate how entanglement develops in the distinct phases of the XX
model.

Finally, our results have shown that the entanglement contents of
the two possible ground states, i.e. the symmetric and the broken-symmetry
state, are different. Therefore, it is of utmost importance to clearly
and explicitly verify which state one is using in any entanglement
analysis made for the XY and related models in order to avoid any
possible confusion.

\textit{Note:}
Upon finishing this manuscript we became aware of a work where the effects
of symmetry breaking on the concurrence are also addressed \cite{Osterloh}, 
and a detailed study of the concurrence and the entanglement between
one site and the rest of the chain (one-tangle) \cite{Verrucchi}.

\begin{acknowledgments}
We acknowledge a couple of email exchanges between A. Osterloh and
one of us (TRO), which helped us to understand some statements
made in Ref. \cite{Osterloh}. We thank Amir O. Caldeira for clarifying
discussions about ME and QPT. We acknowledge support from FAPESP through 
grants  2005/00155-8 (TRO),
 2005/01441-4 (GR), and 04/14605-2 (MCO), and from CNPq through grants 305971/2004-2
(EM), 303918/2005-5 (MCO). This work is also partially supported by the Millennium
Institute for Quantum Information. 
\end{acknowledgments}

\end{document}